\documentclass[%
reprint,
superscriptaddress,
 amsmath,amssymb,
 aps,
 onecolumn,
floatfix,
]{revtex4}

\usepackage{graphicx}
\usepackage{dcolumn}
\usepackage{bm}

\usepackage{gensymb}
\usepackage{color}

\begin{document}

\preprint{APS/123-QED}

\title{$q$-independent  slow-dynamics in atomic and molecular systems}
\author{Philip H. Handle}
\affiliation{Department of Physics, Sapienza – University of Rome, Piazzale Aldo Moro 5, I-00185 Roma, Italy}
\author{Lorenzo Rovigatti}
\affiliation{Department of Physics, Sapienza – University of Rome, Piazzale Aldo Moro 5, I-00185 Roma, Italy}
\affiliation{CNR-ISC, UoS Sapienza, Piazzale Aldo Moro 5, I-00185 Roma, Italy}
\author{Francesco Sciortino}%
\affiliation{Department of Physics, Sapienza – University of Rome, Piazzale Aldo Moro 5, I-00185 Roma, Italy}

\date{\today}

\begin{abstract}
Investigating million-atom systems for very long simulation times, 
we demonstrate that the collective density-density correlation time ($\tau_{\alpha}$) in  simulated supercooled water and silica  becomes wavevector  independent ($q^0$) when the  probing wavelength is several times larger than the interparticle distance.   
The  $q$-independence of the collective density-density correlation functions, a feature clearly observed in light-scattering studies of some soft-matter systems, is thus a genuine feature of many (but not all) slow-dynamics systems, either atomic, molecular or colloidal.  
Indeed, we show that when the dynamics of the density fluctuations is due to particle-type diffusion, as in the case of the Lennard Jones binary mixture model, the $q^0$ regime does not set in and the relaxation time continues to scale as $\tau_{\alpha} \sim q^{-2}$ even at small $q$. 

\end{abstract}

\pacs{Valid PACS appear here}
\maketitle

The wave-vector ($q$) dependence of the dynamics of atomic, molecular, and colloidal systems close to dynamic arrest has been
the focus of intense research~\cite{mezei1987neutron,van1991dynamic,PhysRevE524134,PhysRevE64041503,
sciortino1997supercooled,tolle2001neutron,torre2001acoustic,eckert2002re,bandyopadhyay2004evolution,
bhattacharyya2008facilitation,roldan2017connectivity}.
In particular, the $q$-region that corresponds to the nearest neighbour distance in glass- and gel-forming 
systems has revealed a series of interesting phenomena~\cite{binder2011glassy,donth2013glass}:
(1) a two-step relaxation for both self and collective density correlation functions,
indicating a faster intra-cage motion and a slower structural relaxation ($\alpha$-relaxation), respectively; 
(2) a significant stretching of the $\alpha$-relaxation, which originates from the coupling between distinct modes~\cite{gotze2008complex,balucani1995dynamics}; 
(3) a deviation from the diffusive $q^{-2}$ behavior of the self correlation time; 
(4) oscillations in the $q$-dependence of the collective relaxation time, often in phase with the oscillation of the structure factor;
(5) a faster decay of the self dynamics compared to the collective one, indicating that relative particle displacements play an important role
in the decorrelation of the system.

The region at very small $q$,
where the wavelength is much larger than the inter-particle distance, has also been thoroughly
characterized.  Here, conservation laws in one-component systems determine a three-mode decay of the collective correlation functions~\cite{hansen2013theory}: two modes associated with damped sound waves (the Brillouin peaks) and one to the damped decay of the heat diffusion (the Rayleigh peak). In all three cases, the damping time follows a  $q^{-2}$ dependence. In the case of glasses (where the $\alpha$-relaxation time is longer than the
experimental observation), a clear cross-over has been identified between the region where the system can be considered an elastic continuum and the region where  
an excess of vibrational states~\cite{schirmacher1998harmonic,monaco2009breakdown} is superimposed to the Debye density of states~\cite{tanguy2002continuum,wittmer2002vibrations,zaccone2011approximate}.

In some colloidal systems, where the size of the particles provides access to smaller ratios between the wavelength of the probe radiation and the interparticle distance,
a $q$-independent  ($q^0$) relaxation mode has been reported. From the early measurements in  polymer melts in which entanglement induces an effective transient  network~\cite{brochard1983gel,adam1985dynamical,li2010slow}, evidence of a $q^0$ mode
has been presented for  rodlike micelles~\cite{nemoto1995dynamic}, semidilute polymers with bonding agents~\cite{nemoto1996dynamic, michel2000percolation}, telechelic ionomers~\cite{johannsson1995dynamic},  microemulsion droplets in solution with telechelic polymers~\cite{appell} (where the latter provide transient links between distinct droplets). More recently, a $q^0$ mode has been reported for equilibrium gels of DNA tetrafunctional nanostars~\cite{biffi2015equilibrium,nava} in which a short  self-complementary DNA sequence provides a temperature-controllable  link between different particles. This $q^0$ relaxation dynamics has been interpreted as originating from local elasticity fluctuations  propagating  through the system~\cite{brochard1983gel,nava}. A recent simulation study 
of a particle model for vitrimers, binary-mixture networks in which the microscopic dynamics is slaved to a bond-swapping process~\cite{rovigatti2018self}, has also reported a clear $q^0$ dependence for the collective dynamics extending over more than one order of magnitude in $q$.

In this Letter we explore the possibility that a $q^0$ dependence
of the collective relaxation time is much more common than previously thought, being a generic feature of systems with slow dynamics,
including systems composed of atoms and molecules.  The $q^0$ mode can arise in a range of $q$-values intermediate between the hydrodynamic and the nearest neighbour regions when the collective relaxation is not associated with single particle diffusion.
We simulate slow-dynamics systems that are large enough (up to a million atoms) to allow the numerical evaluation of the collective relaxation over wavelengths corresponding to distances up to 50 times larger than the typical nearest-neighbour distance.
Here we mostly focus on supercooled water whose dynamics has been extensively investigated 
experimentally with different scattering techniques~\cite{monaco1999viscoelastic,bencivenga2007high,torre2004structural,taschin2006supercooled,arbe2016dielectric}. 
For water, a comparison between the ultrasound data~\cite{davis1972liquid} and the lowest  $q$ explored with inelastic x-ray scattering~\cite{monaco1999viscoelastic}  suggests that 
a region with $q^0$ dispersion exist just below the experimentally accessible window. 
Using simulations,  we now provide clear evidence that, indeed, there exists a wide range of 
wave vectors in which the collective dynamics is $q$-independent.  We also show that a model
based on the the Mori-Zwanzig formalism properly describes the $q$-independence of the
dynamics.  Finally, 
we  present results for two more systems, both well-known binary models of glass-formers: BKS silica~\cite{van1990force} and 
a binary mixture Lennard Jones (BMLJ)~\cite{kob1995testing}.
In agreement with the explanation we provide, we observe a $q^0$ mode only in  BKS silica. 

{\it Methods:}
For  TIP4P/2005~\cite{abascal2005general} and BKS~\cite{van1990force}  we perform NVT simulations utilizing GROMACS 5.1.4~\cite{vanderspoel05-jcc} 
with a velocity-Verlet integrator and a timestep of $\Delta t=1$~fs.
The temperature is controlled using a  Nos\'{e}-Hoover thermostat~\cite{nose84-mp,hoover85-pra}. The Coulomb interactions are included with a particle-mesh Ewald treatment~\cite{essmann95-jcp} with a Fourier spacing of 0.1~nm.
For both the Lennard-Jones and real space Coulomb interactions the same cut-off $r_\text{\rm cut}$ is used ($r_\text{\rm cut} = 0.9$~nm and 1~nm for TIP4P/2005 and BKS, respectively).  Lennard-Jones interactions beyond $r_\text{\rm cut}$ are taken into account assuming a uniform fluid density.
The TIP4P/2005 system consists of $2.5\cdot 10^{5}$ molecules (1M interacting sites) in a cubic box. We investigate three densities (0.90, 0.95 and 1.00~g/cm$^3$ corresponding to box lengths of 20.26, 19.89, and  19.56~nm, respectively) at two temperatures (240 and 250~K). The molecule constraints are maintained using the LINCS (Linear Constraint Solver) algorithm~\cite{hess08-jctp}.
The BKS system consists of 287712 particles (1/3 Si and 2/3 O) in a cubic box.
We study three densities (2.36, 2.48 and 2.60~g/cm$^3$, corresponding to box lengths of 15.95, 15.69 and 15.45~nm) at 3500~K. As a binary-mixture Lennard-Jones (BMLJ) glass-former, among the different alternatives~\cite{PhysRevA.38.454,roux1989dynamical,wahnstrom1991molecular}, we choose the Kob-Andersen model~\cite{kob1995testing}. We use the RUMD package~\cite{bailey2017rumd} to simulate $10^6$ particles at $\rho^* = 1.2$ and $T^* = 0.466$,
where the $^*$ superscript indicates reduced units: the unit of length is the diameter of the large particles, $\sigma_{AA}$, the unit of mass is the particle mass $m$ and the unit of energy is the interaction strength between large particles, $\epsilon_{AA}$. Differently from Ref.~\cite{kob1995testing}, the unit of time is $\sqrt{m\sigma_{AA}/\epsilon_{AA}}$. The total duration of the simulations is 20 ns for TIP4P/2005 and BKS (20 to 100 times the slowest collective relaxation time in the system) and $6 \times 10^7$~ MD time steps for the BMLJ system. 
The relaxation times and non-ergodicity factors obtained from fitting  the correlation functions of the TIP4P/2005 and BKS systems  in the whole $q$-range are reported in the Supplemental Material (SM).

\begin{figure}[h!]
   \centering
      \includegraphics[width=0.44\textwidth]{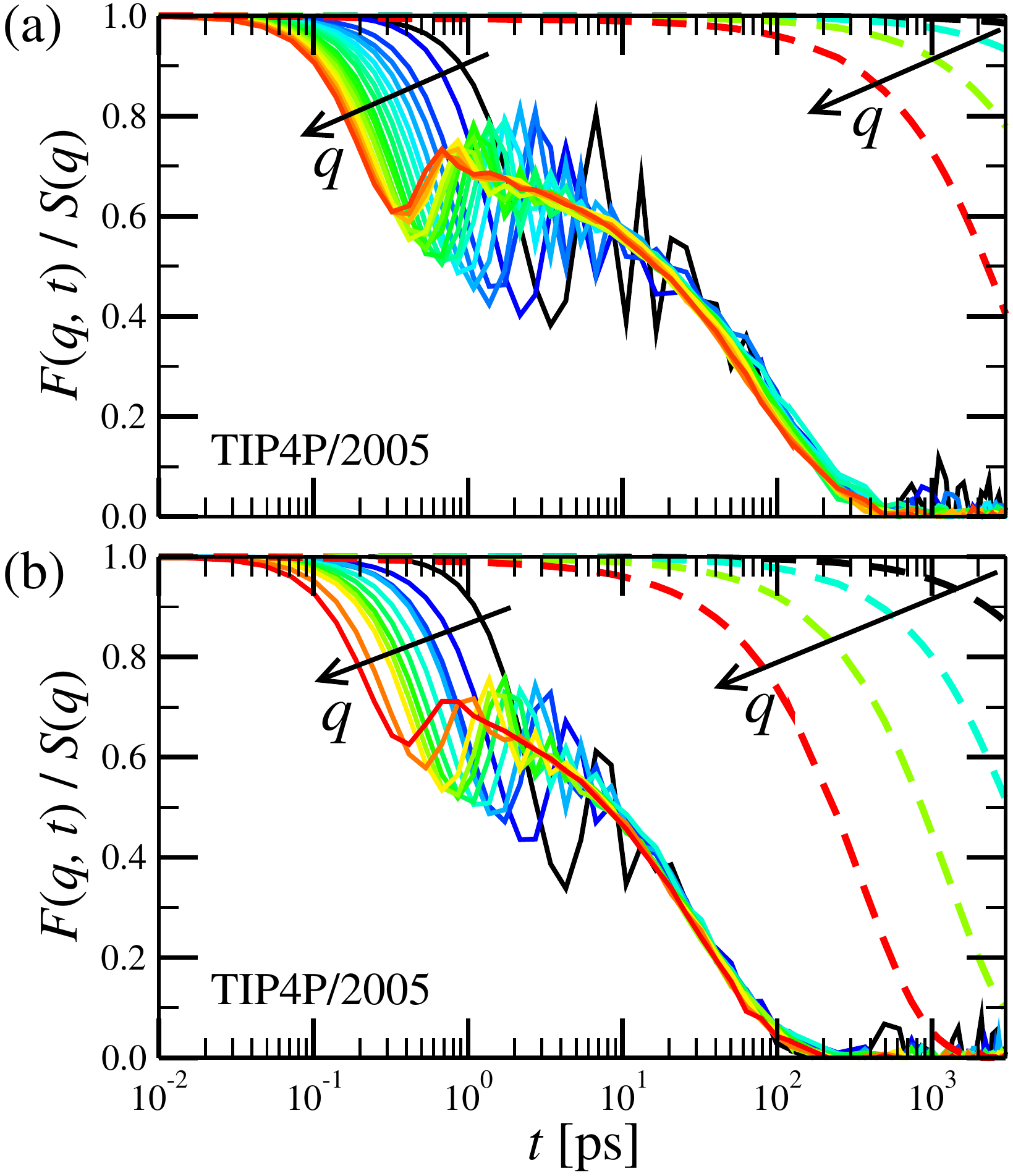} 
   \caption{Decay of the (full lines) collective and (dashed lines) self density fluctuations at (a) $T=240$~K $\rho=0.9$ g/cm$^3$
      and (b) $T=250$~K $\rho=1.0$ g/cm$^3$. In both panels $q$ varies between
   $0.3$ (black lines) and $3.0$ (red lines) nm$^{-1}$.}
   \label{fig:tip4p}
\end{figure}

{\it Results --- TIP4P/2005:}
Fig.~\ref{fig:tip4p} shows the (oxygen-oxygen) collective density correlation functions $F(q, t)$ for several $q$ values in the window $0.3<q<3$ nm$^{-1}$, corresponding approximatively to wavelengths 7 to 70 times larger than the OO nearest neighbour distance (0.28~nm). $F(q, t)$ is evaluated as
\begin{equation}
F(q, t)=  \langle \rho_{\bf q}^*(t) \rho_{\bf q}(0) \rangle, ~~~~  \rho_{\bf q}(t) \equiv \frac{1}{\sqrt{N}} \sum_{i=1}^{N} e^{i {\bf q} \cdot {\bf r}_i(t)}
\end{equation}
where $N$ is the total number of molecules, $ {\bf r}_i$ the position of the $i$-th oxygen atom, and the average is performed over different initial times along the MD trajectory and over all ${\bf q}$ vectors with the same modulus $q$. $F(q, 0)$ coincides with the
structure factor $S(q)$.  It is immediately clear from Fig.~\ref{fig:tip4p} that 
in this $q$ window (extending the window of $q$ explored with 
high-resolution inelastic x-ray scattering measurements~\cite{monaco1999viscoelastic,bencivenga2007high}), all correlation functions decay at long time with the same characteristic time scale. Fig.~\ref{fig:tip4p} also shows the self-correlation function $F_{\rm self}(q, t)$, plotted with dashed lines. In this small-$q$ region the Gaussian approximation holds true, that is,  $F_{\rm self}(q, t)=\exp(-q^2 \left\langle r^2(t) \right\rangle / 6)$, where $\left\langle r^2(t) \right\rangle$ is the mean-squared displacement\cite{hansen2013theory} (comparison not shown). It is important to note that the self correlation functions decay at much longer times compared to $F(q,t)$, indicating that the decorrelation of the collective dynamics over the probed length-scales does not require particle diffusion.

To further support the idea that a single ($q$-independent) characteristic time controls the 
decay of the density fluctuations in this low-$q$ region we compare the numerically calculated correlation functions with predictions based on the Mori-Zwanzig (MZ) formalism. For any system, the normalised collective correlation function  $\phi_q \equiv F(q,t)/S(q)$ formally satisfies the equation~\cite{hansen2013theory}
\begin{equation}
\frac{d^2\phi_q}{dt^2 } +\Omega^2_q \phi_q +\int_0^t M_q(t-t') \frac{d\phi_q}{dt} dt' =0
\end{equation}
where $\Omega^2_q= \frac{k_BT}{M} \frac{q^2}{S(q)}$.

The memory function, which is the autocorrelation of the stochastic force~\cite{hansen2013theory}, and $\Omega^2$ completely specify the
dynamics. Guided by the $q$ independence of $\tau_{\alpha}$ we approximate $M_q(t)$ as

\begin{equation}
M_q(t)=\gamma_0 q^{2}\delta(t)+ M^s_q(t),~~ M^s_q(t)  \equiv A_s q^2 \exp{[-(t/\tau_s)^\beta]}
\end{equation}
\noindent
where $\gamma_0 q^{2}\delta(t)$ describes the damping associated with the  
fast microscopic dynamics and $M^s_q(t)$ models the structural relaxation. This function
is characterised by the three  
 parameters   $A_s$, $\tau_s$ and $\beta$, defining respectively the
amplitude and the time-dependence of $M^s_q(t)$.  All four parameters ($\gamma_0$, $A_s$, $\tau_s$ and $\beta$)
are in principle functions of $q$.
Consequently, 
\begin{equation}
\frac{d^2\phi_q}{dt^2 } +\Omega^2_q \phi_q + \gamma_0 q^2 \frac{d\phi_q}{dt}  +\int_0^t 
M^s_q(t-t') 
\frac{d\phi_q}{dt} dt' =0.
\label{eq:modelM}
\end{equation}
In deriving Eq.\eqref{eq:modelM} we  have neglected  the coupling between density and temperature fluctuations and
the thermal diffusion contribution which is (in the hydrodynamic limit) proportional
to $C_v/C_p-1$~\cite{stanley1971phase}. For water this is a reasonable approximation, since at the temperature of maximum density $C_v=C_p$~\cite{stanley1971phase}.

\begin{figure}[h!] 
\centering
\includegraphics[width=0.44\textwidth]{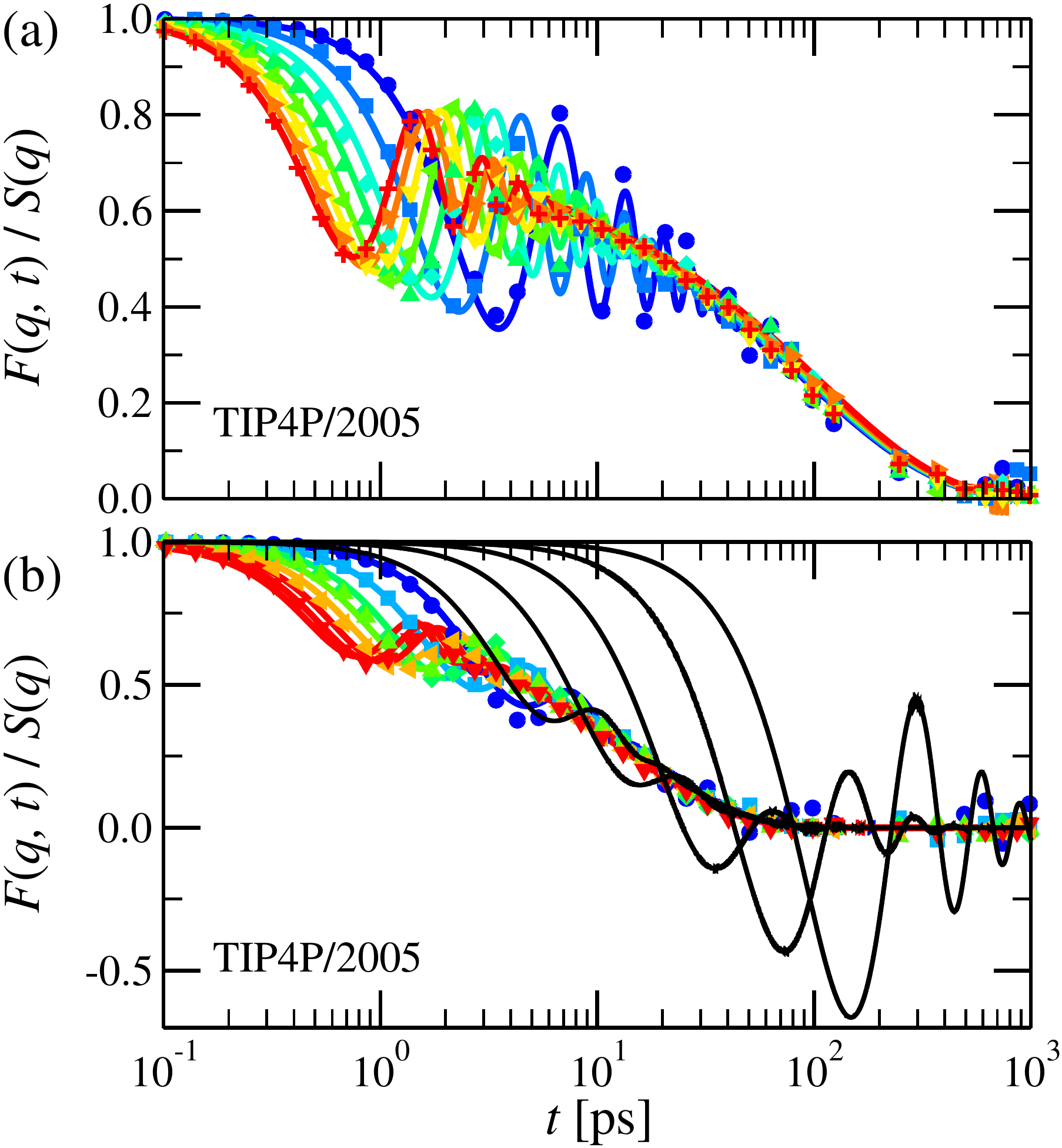} 
\caption{Simulation (symbols) and theoretical (lines) results for (a) $\rho=0.90 $ g/cm$^3$ and $T=240 $K and (b) for $\rho=1.0 $ g/cm$^3$ and $T=250 $K, in the same $q$ window as in Fig.~\ref{fig:tip4p}.
For the prediction of Eq.~\eqref{eq:modelM}, in (a) $\gamma_0=0.4$, $M_{q}^s(t)=7.3 q^2 \exp [-(t/15)^{0.5} ]$ 
while in (b) $\gamma_0=0.3$,  $M_{q}^s(t)=10 q^2\exp[-(t/1.05)^{0.5}]$ 
(with $t$ in ps, $q$ in nm$^{-1}$, $\gamma_0$ in ps$^{-1}$nm$^{2}$ and  $M_q$ in ps$^{-2}$). In (b) the black lines show predictions for very small $q$ vectors (from right to left $q_n=0.0016 \times 2^{n-1}$  nm$^{-1}$ with $n=1,2,3,4,5$), to highlight the cross-over from structural relaxation to hydrodynamics.}
\label{fig:modelM}
\end{figure}

We solve the time dependence of Eq.~\eqref{eq:modelM} numerically  (in Fourier space), searching for the values 
of  the parameters which minimize the differences  with 
the  correlation functions evaluated from the simulated trajectories. We find that, in the range $0.3<q<3$ nm$^{-1}$, it is possible to represent 
all correlation functions in the entire
time window by using $q$-independent  values for $\gamma_0$, $A_s$, $\tau_s$ and $\beta$.   Fig.~\ref{fig:modelM} compares the $q$-independent 
model predictions  with the numerical data for two different state points.
Such excellent agreement  provides strong evidence that a collection of local relaxation processes (as indicated
by the value $\beta=0.5$) fully describes the $q$-independent decay of the collective dynamics at small $q$. 
The prediction of Eq.~\eqref{eq:modelM} for $q$ smaller than the ones numerically accessed (black lines in Fig.~\ref{fig:modelM}(b))  visually demonstrate the cross-over to the hydrodynamic limit (having neglected the
$T$-fluctuations, the limit coincides with the damped harmonic oscillator
model with adiabatic sound speed $v_s= \Omega_q/q$~\cite{hansen2013theory,balucani1995dynamics}).
In general, hydrodynamics sets in when the time scale of the non-structural  component  becomes
lower than $\tau_{\alpha}$. Selecting the period of the sound wave as the typical time, 
the wave-vector below which the relaxation time loses its $q^0$ character  is reached when $q \ll 2 \pi / (v_s \tau_\alpha)$. Therefore, upon approaching the glass transition, as $\tau_\alpha$ becomes longer and longer, the region of $q$  in which a $q^0$ mode is expected becomes larger and larger, provided that $S(q)$ does not vary significantly.

\begin{figure}[h!]
   \centering
      \includegraphics[width=0.44\textwidth]{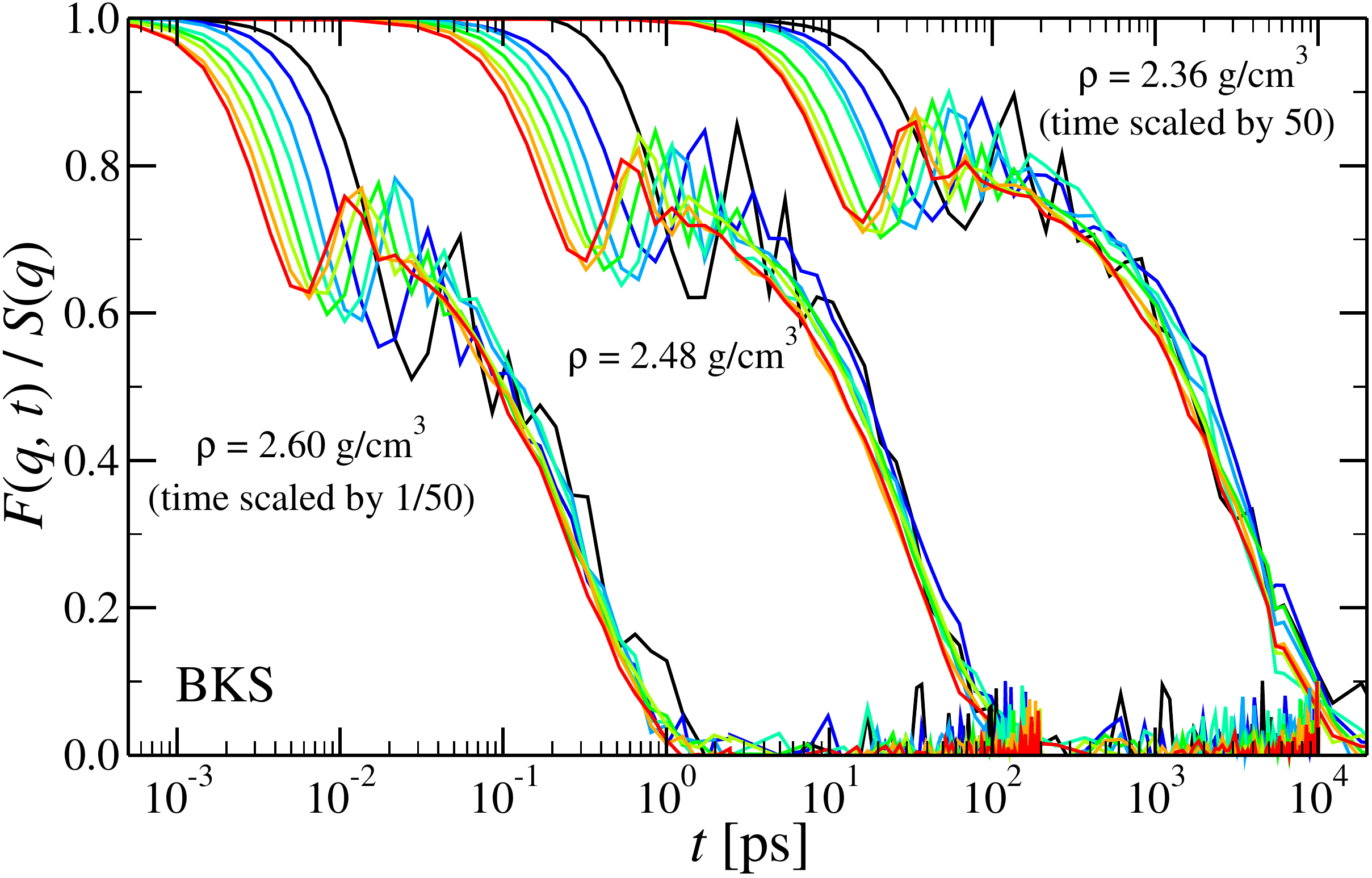} 
   \caption{Decay of the collective number density fluctuations in BKS silica at $T=3500$ K for 
   different densities.  In all plots, $0.4<q<2$ nm$^{-1}$ (from black to red). Note that, for the sake of clarity, the time associated with the $\rho=2.60$ g/cm$^3$ and $\rho=2.63$ g/cm$^3$ data has been multiplied by 1/50 and 50, respectively. As a reference, the Si-Si distance is $\approx 0.31$~nm.}
   \label{fig:bks}
\end{figure}

\begin{figure}[h!]
   \centering
      \includegraphics[width=0.44\textwidth]{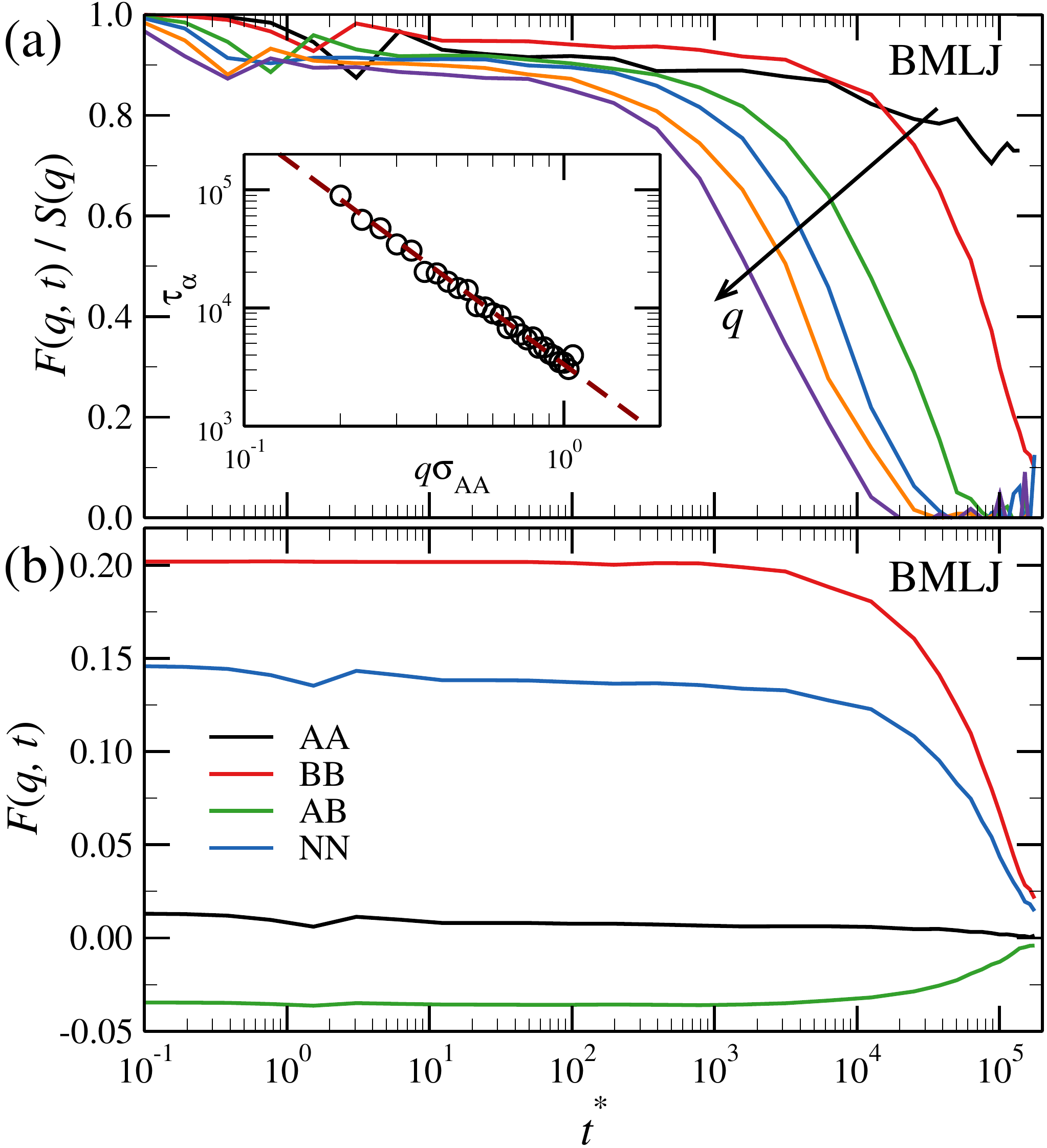}
   \caption{Density-density collective correlation functions 
for BMLJ at $T^*=0.466$ and $\rho^*=1.2$. (a) $F(q, t)$ calculated for all particles, normalised by the structure factor $S(q)$. From right to left $q_n \sigma_{\rm AA}=0.06676 \times n$ with $n=3,6,12,18,24,30$. The inset shows the corresponding relaxation time $\tau_\alpha$ as a function of $q$. The dashed line, which has a slope given by $q^{-2}$, is shown for comparison. (b) The non-normalised collective correlation functions calculated for $q\sigma_{\rm AA} = 0.4$ between AA, AB, BB and all particles (NN).}
   \label{fig:bmljsmallq}
\end{figure}

{\it Results --- BKS and BMLJ:}
We next investigate two of the most commonly-studied binary mixture glass-former models.
BKS silica~\cite{van1990force}  generates a structure in which the (positively charged) Si atoms form a prevalently 
tetrafunctional open network in which the (negatively charged) O atoms mediate the Si-Si bonds. As expected for charged systems, there are essentially no large wavelength concentration fluctuations~\cite{stillinger1968ion}. By contrast, BMLJ~\cite{kob1995testing} generates a dense structure in which each particle is surrounded by particles of both species. BMLJ particles are electrically neutral and   concentration fluctuations do not vanish in the small $q$ limit. This crucial difference is evident in the Bhatia-Thornton concentration-concentration structure factors~\cite{bhatia1970structural} (reported in the SM).
In the small-$q$ region explored in the simulation, the decay of the collective total-density fluctuations in
the two binary systems is completely different.  While 
BKS (Fig.~\ref{fig:bks}) behaves as TIP4P/2005 water and displays an almost $q^0$ dependence for all three investigated densities, the same function in  BMLJ 
manifests a strong $q^{-2}$ dependence (see main panel and inset of Fig.~\ref{fig:bmljsmallq}(a)). We also show the (non-normalised) partial correlation functions for $q\sigma_{\rm AA} = 0.4$ in Figure~4(b). All the correlation functions decay similarly in the whole $q$-range considered, and the associated relaxation times always display a $q^{-2}$ dependence.  Thus, in contrast to the one-component water case and the two-components silica case,  the small-$q$ $\alpha$-relaxation process in BMLJ does not acquire a $q$-independent behaviour.

{\it Discussion and conclusions:}
The reported numerical investigation of the low-$q$ collective dynamics in 
 atomic and molecular systems exhibiting slow dynamics provides indisputable numerical evidence of the existence of a large region of $q$, corresponding to wavelengths larger than
the inter-particle distance, in which the relaxation time is $q$-independent.  These findings demonstrate 
that a $q^0$ mode is not a prerogative of soft-matter systems, but it is a genuine feature of many slow-dynamics systems, either atomic, molecular or colloidal. 
The $q$-independence of $F(q,t)$ can be equivalently expressed as a $q$-independence of the memory function within the MZ formalism.

Interestingly, from the  point of view of the mode coupling theory (MCT), which neglects crystallisation, 
the difference between the $q$-dependence of the memory functions associated with the structural relaxation of disordered one-component systems  and binary mixtures can be traced back to the conservation of the overall momentum of the system~\cite{gotze2008complex,weysser2010structural}. In one component MCT, consistent with what we have
found,  $M_q$ is predicted to be $q$ independent at small $q$.
In the case of binary mixtures,
 momentum  conservation  is not observed by each component, since only the sum of all partial momenta is conserved. As a result, the MCT relaxation times  might display a $q^{-2}$ dependence~\cite{gotze2008complex,weysser2010structural}. Therefore, within this framework, our results seem to indicate that binary mixtures for which the large-wavelength concentration fluctuations are suppressed  behave, at large length-scales, as effective one-component systems. A comparison between the kinetic and thermodynamic contributions to the interdiffusion constant of the BMLJ and BKS systems clearly demonstrate the presence of strong cross correlations only in the latter (see SM)~\cite{horbach2007self,zausch2009statics,hansen2013theory}.
These predictions have never been carefully tested due to the numerical difficulties of simulating very large systems over very long periods of time. Our results will  definitively stimulate novel theoretical analysis to properly frame the $q^0$ phenomenon. In particular, it will be interesting to understand in more depth under which conditions binary mixtures do or do not exhibit a $q^0$ mode.  The reported results will also stimulate new small-angle neutron and x-ray experiments, where a confirmation of the predicted behaviour can possibly be achieved.
The  results presently available for water~\cite{monaco1999viscoelastic,bencivenga2007high} (reaching a lowest $q$ of  1 nm$^{-1}$) are consistent with the numerical findings discussed here.

Although the memory function formalism provides a deep formal understanding
of the $q$ independence, it does not offer a physical picture of the 
local processes responsible for probing all different available microstates at small length scales.
Different systems will relax locally in different ways, possibly via rototranslational motion (as proposed for water~\cite{saito2018crucial}), bond-swapping or bond flickering in networks.
Despite these differences,  the effect of such local decorrelations on the small $q$ dynamics  is system independent. The
local changes of elasticity brought in by these local events possibly propagate via the vibrational modes of the system~\cite{nava}
resulting in the decay of the collective density ($N/V$) fluctuations at large distances  controlled by changes not in the number of particles $N$ but in the volume $V$.

Finally, we  observe that in the $q^0$ regime equilibration in systems of very large size  can be achieved only if simulations  can be run for times longer than $\tau_{\alpha}$ (since $\tau_{\alpha}$ does not grow with decreasing $q$).
This is not the case for BMLJ (and possibly other binary Lennard-Jones glass formers~\cite{PhysRevA.38.454,roux1989dynamical,wahnstrom1991molecular}), for which doubling the system size requires
four times longer equilibration times, making it essentially impossible to generate
very large  structurally and compositionally equilibrated configurations.  Indeed, in these cases the diffusion of individual particles is required to relax frozen-in long wavelength  concentration fluctuations. The observed $q^0$ mode in binary systems, shown here for BKS and previously for binary-mixture vitrimers~\cite{rovigatti2018self},  can possibly originate from the suppression of the small wavevector (large wavelength) concentration fluctuations.  For BKS, this effect stems from the  electrostatic nature of the  Si and O interactions~\cite{stillinger1968ion}. By contrast, in the vitrimer system~\cite{rovigatti2018self} it is the precise stoichiometry of the mixture and the bonding mechanism, which can act only between unlike particles, that prevents the occurrence of small-$q$ concentration fluctuations. 

\begin{acknowledgments} 
We thank G. Monaco and J. F. Douglas for discussions, and J. Russo and C. Scherfler  for comments on the manuscript. PHH acknowledges financial support from the Austrian Science Fund (FWF Erwin Schr\"{o}dinger Fellowship J3811 N34). LR acknowledges support from the European Research Council (ERC Consolidator Grant 681597, MIMIC).
\end{acknowledgments}

\bibliography{binary}

\section{Supplemental Material}

\subsection{Collective relaxation times, non-ergodicity and Debye-Waller factors}

We fit the density-density collective correlation functions with the functional form

$$
F_c(t)=(1 - f)\exp(-(t/\tau_f)^2)\cos(\omega t)+f\exp(-(t/\tau_\alpha)^\beta) 
$$

\noindent
where $f$, $\tau_f$, $\omega$, $\tau_\alpha$ and $\beta$ are fitting parameters. 

\begin{figure}[h!]
\centering
\includegraphics[width=0.44\textwidth]{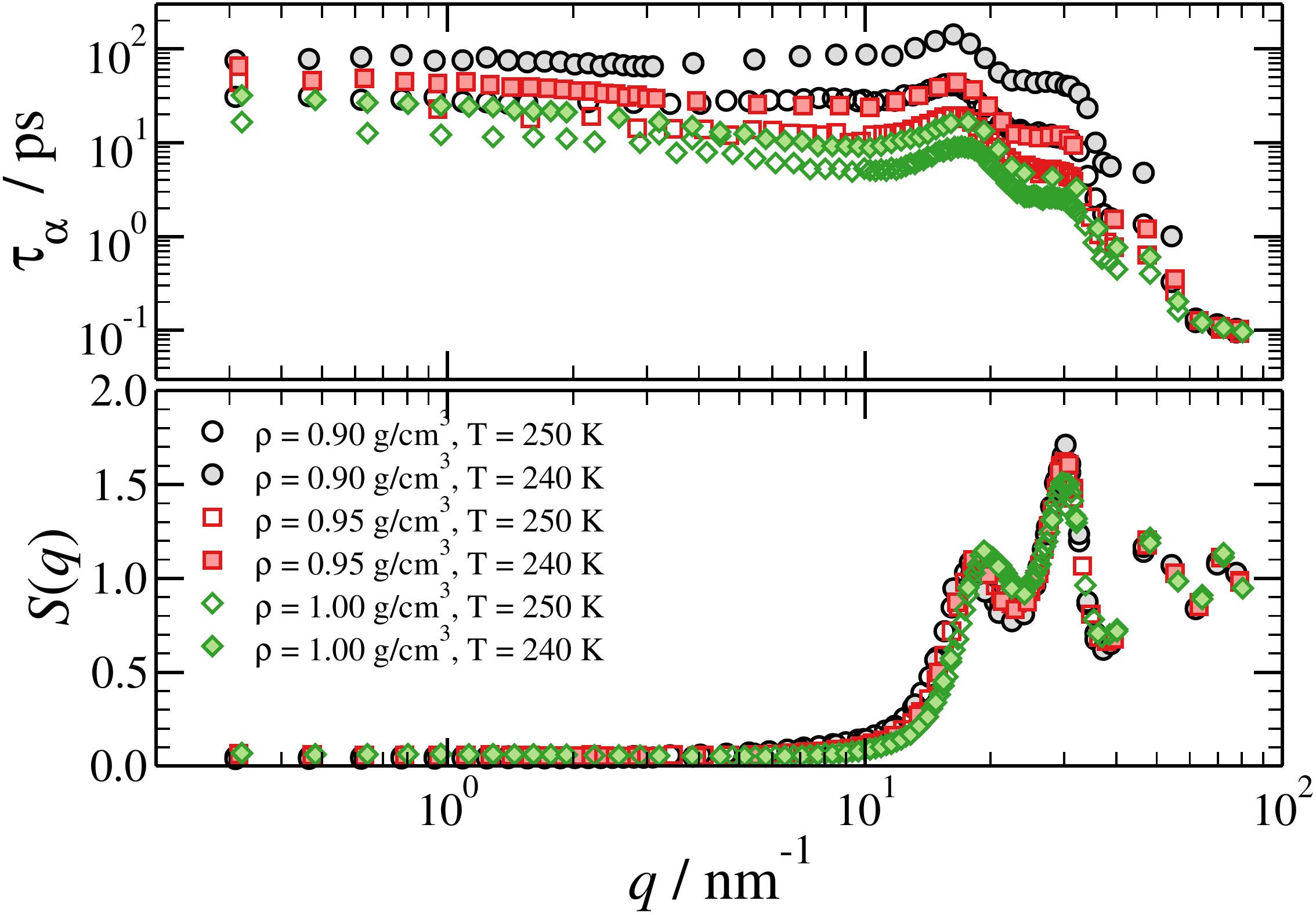}
\includegraphics[width=0.44\textwidth]{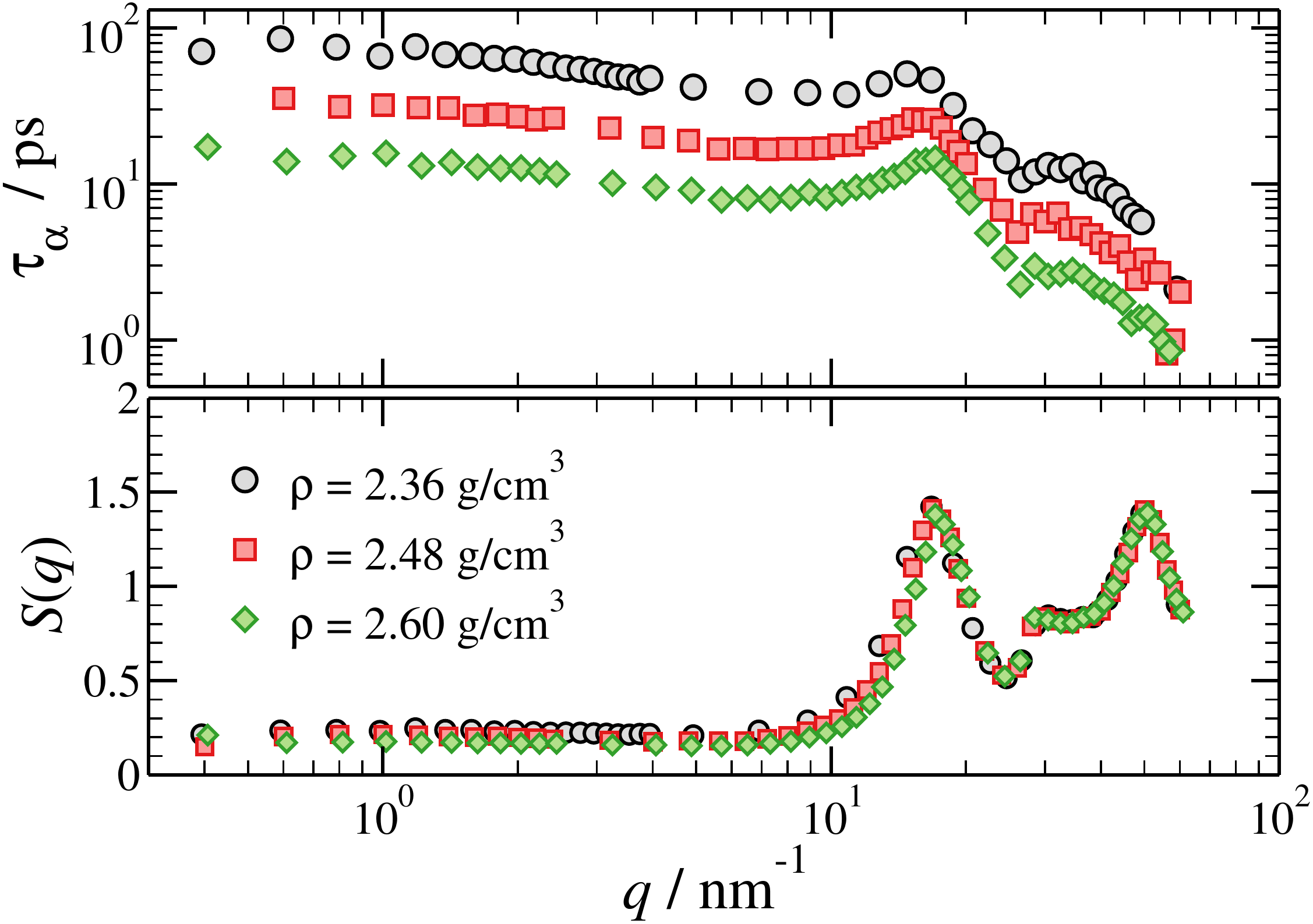}
\caption{Relaxation time $\tau_\alpha$ (upper panels) and structure factor $S(q)$ (lower panels) as functions of the wavevector $q$ for the (left) TIP4P/2005 and (right) BKS systems.}
\label{fig:tau}
\end{figure}

Figure~\ref{fig:tau} shows the collective relaxation time $\tau_\alpha$ for all the investigated TIP4P/2005 and BKS systems.

\begin{figure}[h!]
\centering
\includegraphics[width=0.44\textwidth]{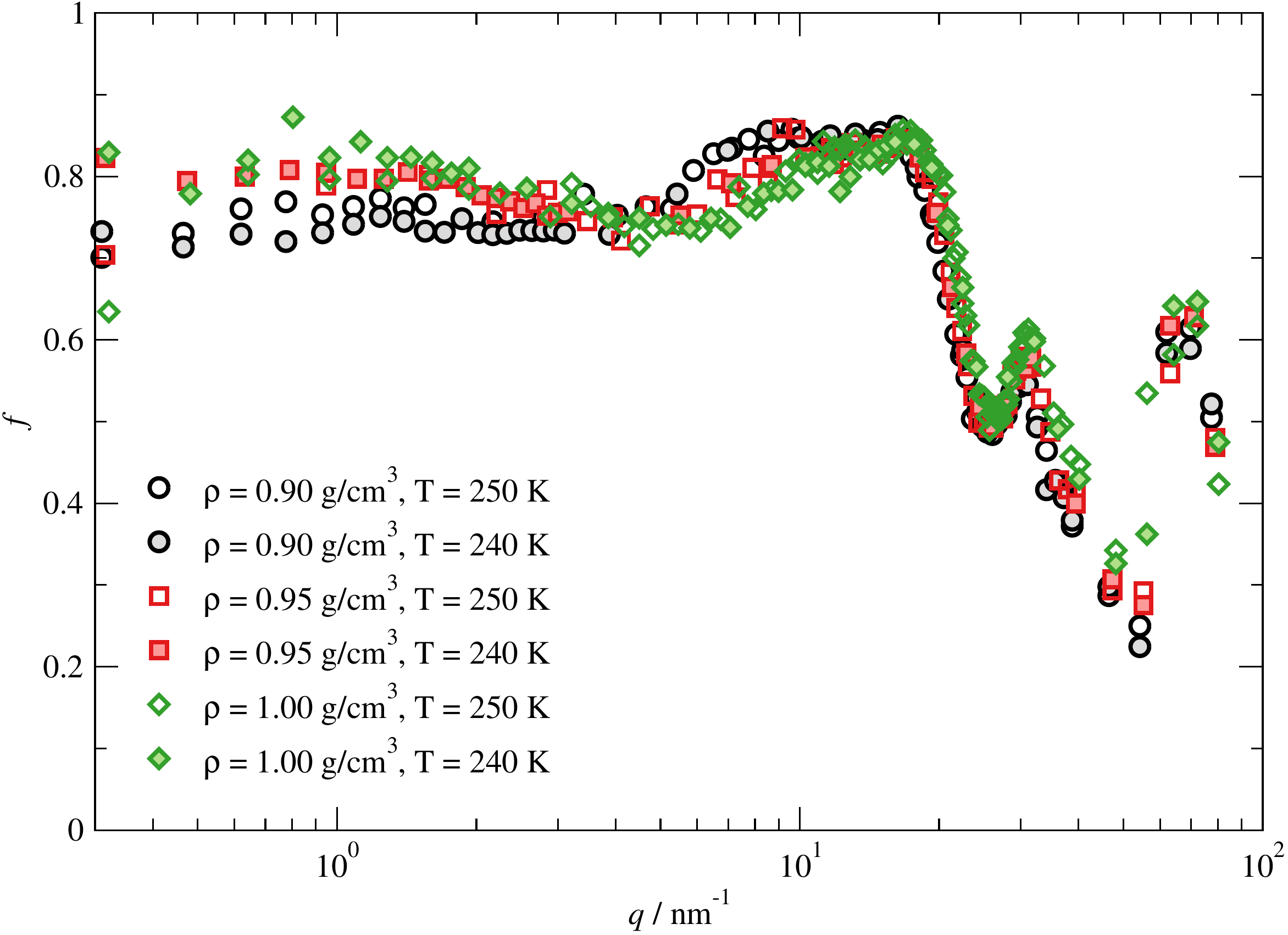}
\includegraphics[width=0.44\textwidth]{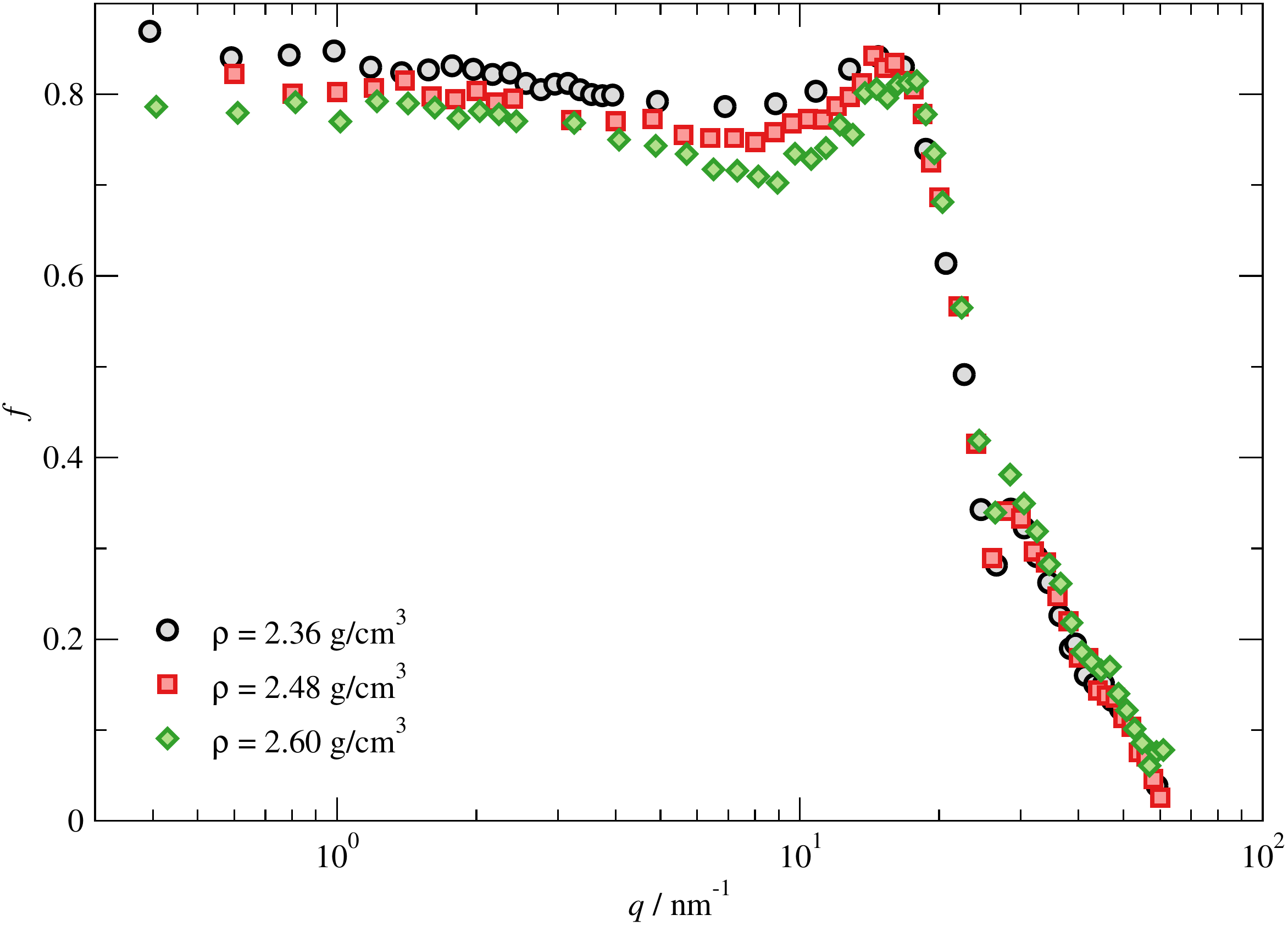}
\caption{Non-ergodicity factor $f$ as a function of $q$ for the (left) TIP4P/2005 and (right) BKS systems.}
\label{fig:f}
\end{figure}

Figure~\ref{fig:f} shows the non-ergodicity factor $f$ for all the investigated TIP4P/2005 and BKS systems.

\begin{figure}[h!]
\centering
\includegraphics[width=0.44\textwidth]{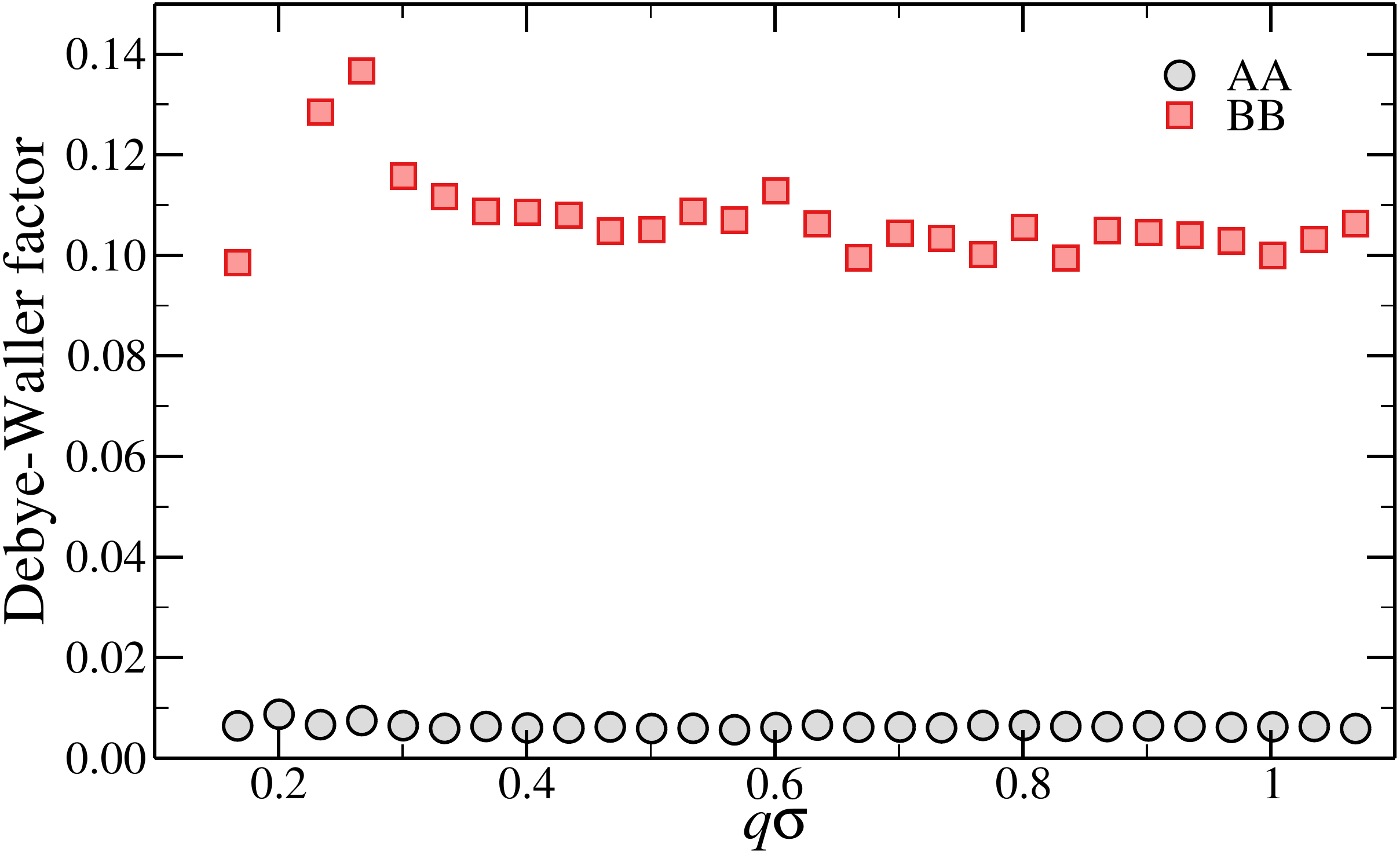}
\includegraphics[width=0.44\textwidth]{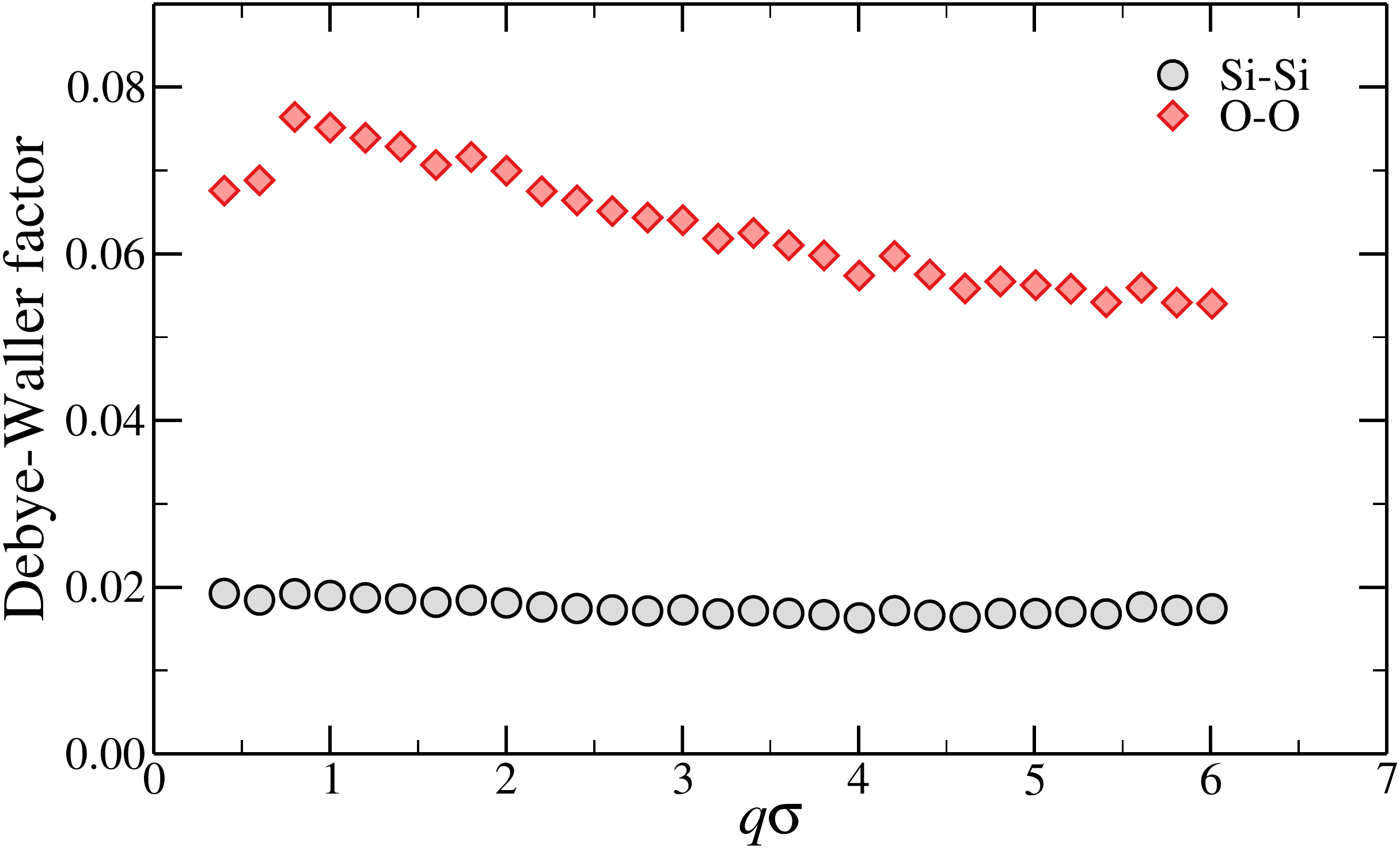}
\caption{Debye-Waller factors for the (left) BMLJ and (right) $\rho = 2.48$ g/cm$^3$ BKS systems.}
\label{fig:dw}
\end{figure}

Figure~\ref{fig:dw} shows the partial Debye-Waller DW factors for the BMLJ and one BKS system. In the BMLJ system the DW factor differs greatly between the two species. By contrast, in BKS silicon and oxygen behaves more similarly, as their associated DW factors have similar values. We note that the values reported here are in agreement with the ones presented in Ref.~\cite{PhysRevE.64.041503}.

\subsection{Interdiffusion and concentration fluctuations}

\begin{figure}[h!]
\centering
\includegraphics[width=0.6\textwidth]{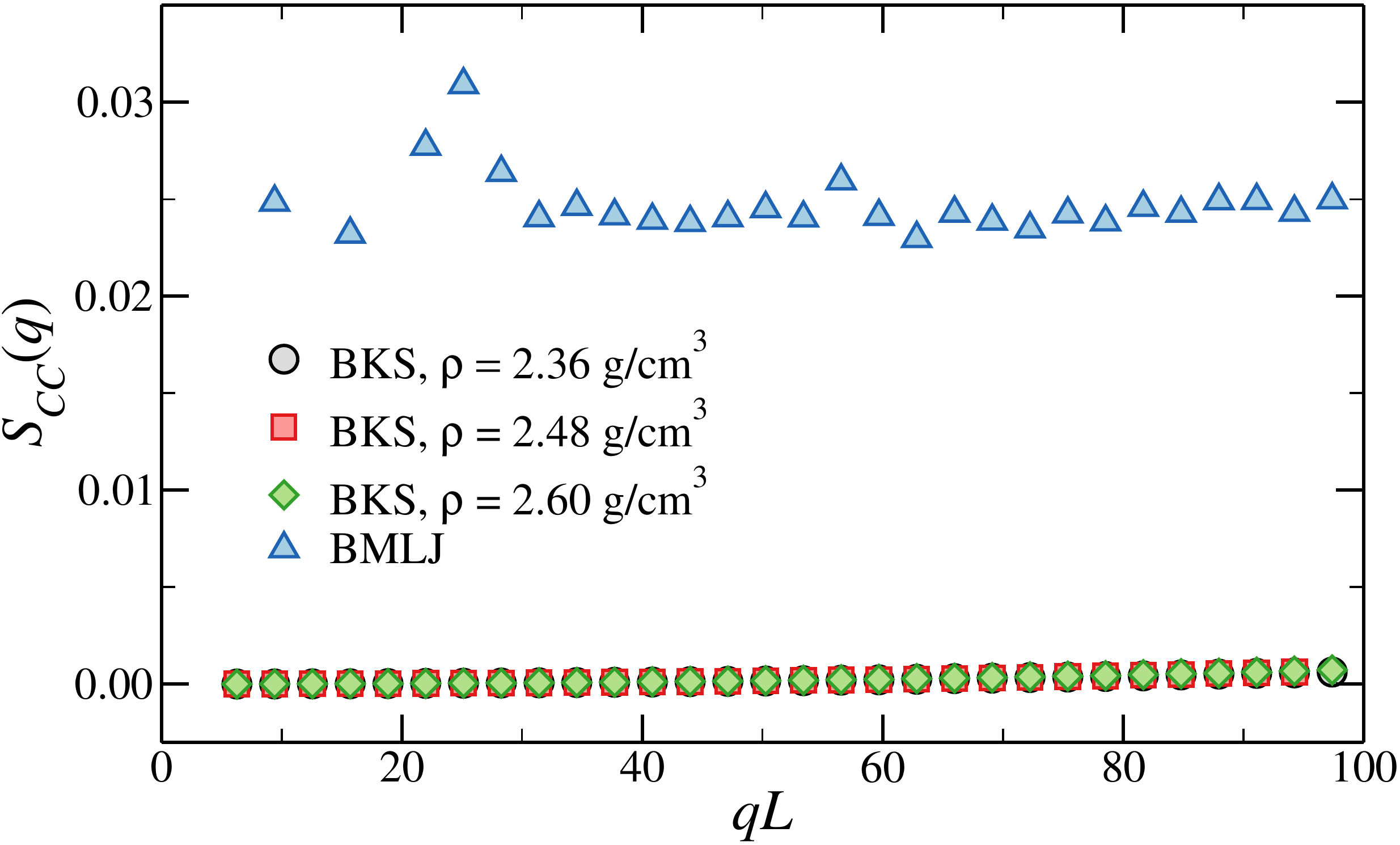}
\caption{The small-$q$ part of the Bhatia-Thornton concentration-concentration structure factor for the BKS and BMLJ systems~\cite{bhatia1970structural}. The values of the $S_{CC}(q)$ calculated for BKS for the smallest wavevectors are of the order of $10^{-6}$, roughly four orders of magnitude smaller than the corresponding BMLJ values.}
\label{fig:sqCC}
\end{figure}

For a generic binary mixture composed of particles of species $A$ and $B$, the extent of the structural cross-species correlations can be estimated by computing the Bhatia-Thornton concentration-concentration structure factor $S_{CC}(q)$~\cite{bhatia1970structural}, which is defined as

$$
S_{CC}(q) = (1 - x)^2 S_{AA}(q) + x^2 S_{BB}(q) - 2 x (1 - x) S_{AB}(q),
$$

\noindent
where $x$ is the fraction of particles of type $A$ and $S_{AA}(q)$, $S_{BB}(q)$ and $S_{AB}(q)$ are the partial structure factors. Figure~\ref{fig:sqCC} shows the small-$q$ limit of $S_{CC}(q)$ for the BKS (for which $A$ and $B$ are silicon and oxygen, respectively) and BMLJ systems. The curves clearly show that the BKS system exhibits exceedingly small large-wavelength concentration fluctuations with respect to the BMLJ system.

\begin{figure}[h!]
\centering
\includegraphics[width=0.44\textwidth]{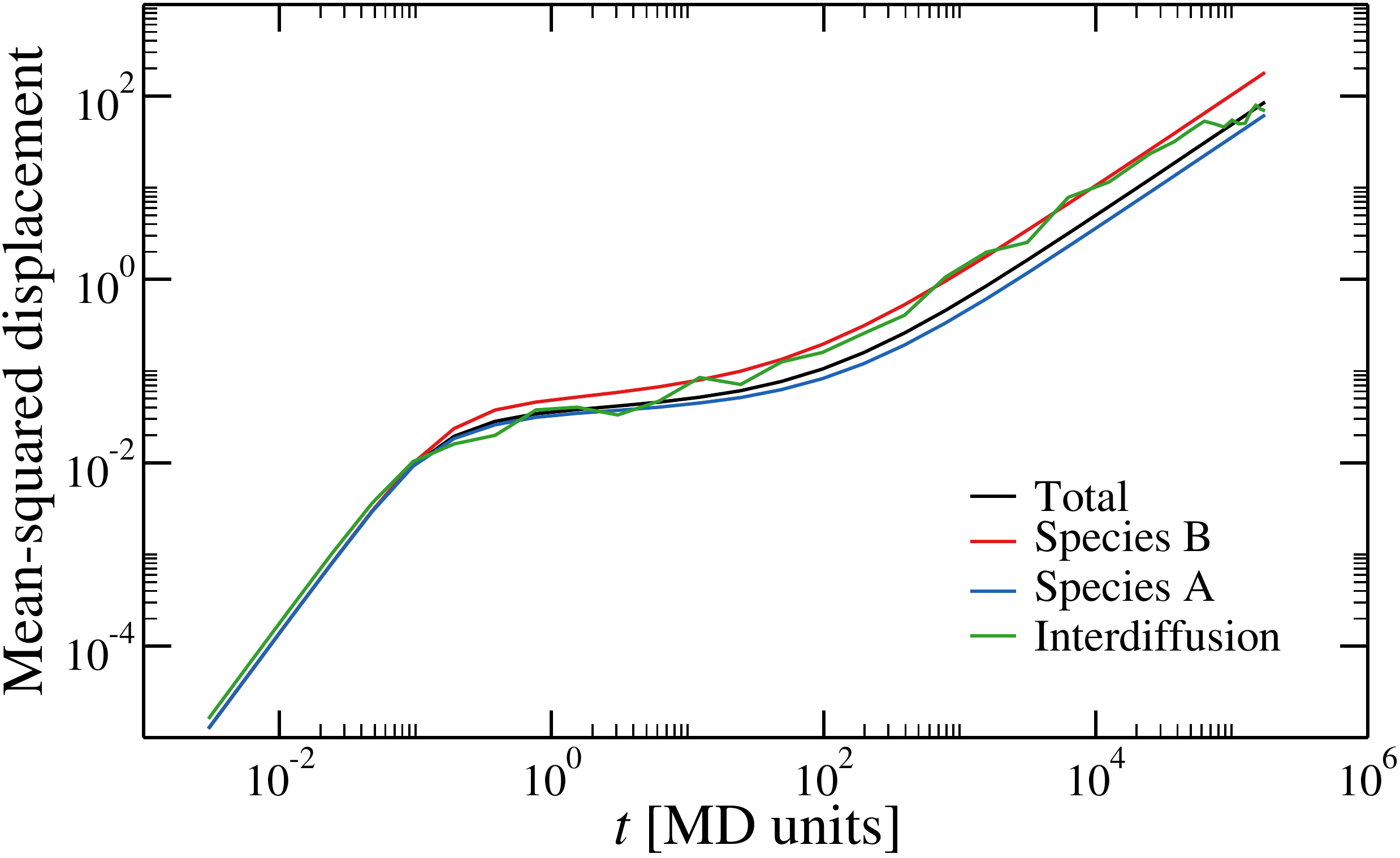}
\includegraphics[width=0.44\textwidth]{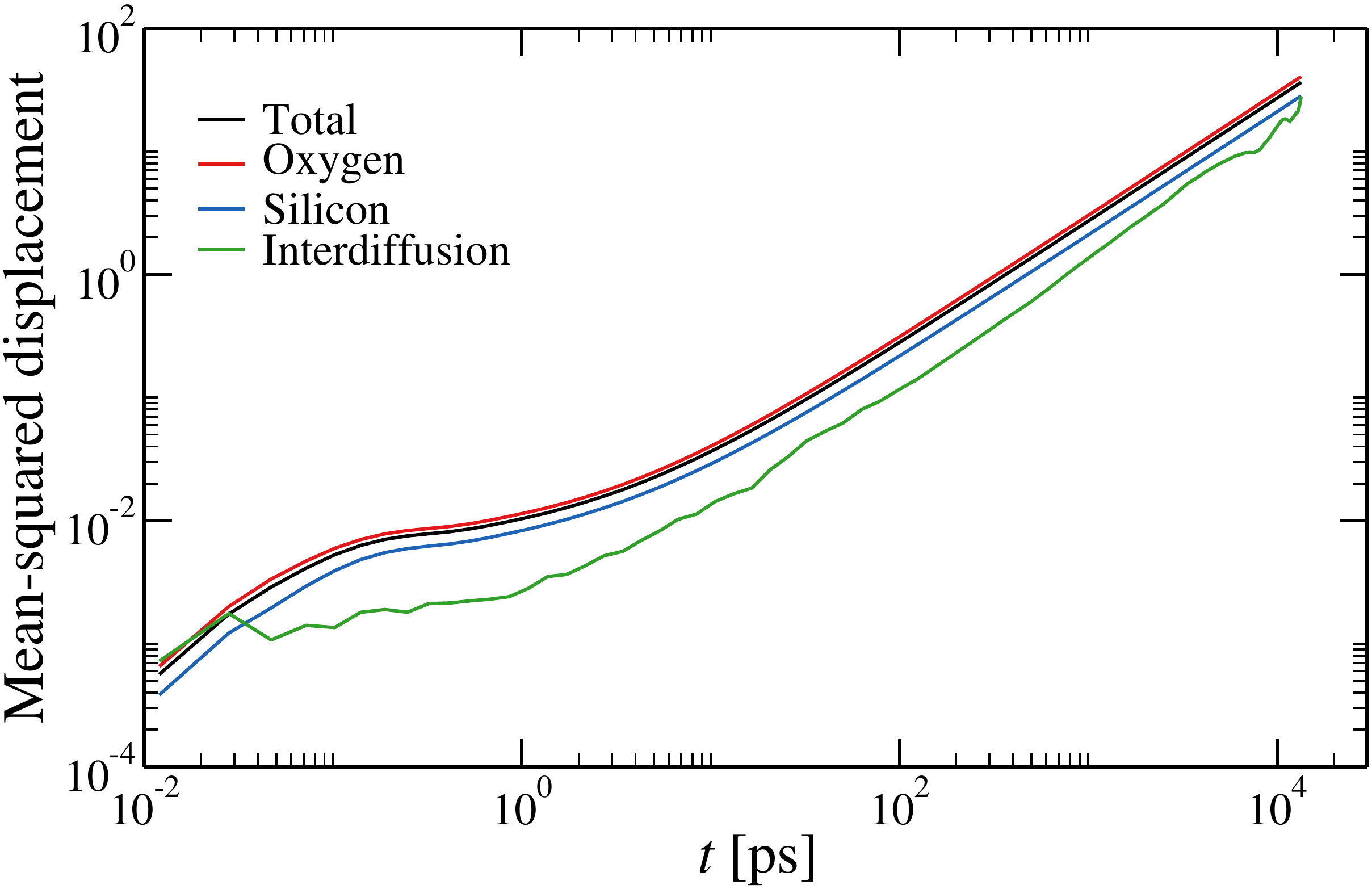}
\caption{Mean-squared displacements for the (left) BMLJ and (right) BKS ($\rho = 2.60$ g/cm$^3$) systems. In addition to the total and partial mean-squared displacements, we also show the MSD of the centre of mass of particles of species A for the BMLJ and of the Silicon for the BKS (green lines).}
\label{fig:interdiffusion}
\end{figure}

The collective transport of mass driven by concentration fluctuations can be quantified by the so-called interdiffusion~\cite{hansen2013theory,horbach2007self}. The associated transport coefficient is the interdiffusion constant, which can be written as~\cite{horbach2007self,zausch2009statics}

$$
D_{AB} = \Phi \Lambda
$$

\noindent
where $\Phi$ is a thermodynamic factor and $\Lambda$ is a kinetic factor also known as the interdiffusion Onsager coefficient. The former is defined as

$$
\Phi = \frac{x(1-x)}{S_{CC}(q = 0)},
$$

\noindent
while the latter is related to the long-time behaviour of the mean-squared displacement of the centre of mass of one of the two species. In particular, let $\vec{R}_A(t)$ be the position of the centre of mass of species $A$ at time $t$ and

$$
g_{\rm int}(t) \equiv \left( 1 + \frac{m_A x}{m_B (1 - x)}\right)^2 N x (1 - x) \left\langle \left[ \vec{R}_A(t) - \vec{R}_A(0) \right]^2 \right\rangle
$$

\noindent
its associated mean-squared displacement, weighted by a factor that takes into account the mass of particles of species $A$ and $B$ ($m_A$ and $m_B$, respectively), their concentration and the total number of particles, $N$. Figure~\ref{fig:interdiffusion} shows $g_{\rm int}(t)$ together with the total and partial mean-squared displacements for the BMLJ and one BKS system. In terms of $g_{\rm int}(t)$, the Onsager coefficient is given by

$$
\Lambda = \lim_{t \to \infty} \frac{g_{\rm int}(t)}{6 t}.
$$

We have already analysed the difference between the static concentration-concentration correlations as estimated by the small-$q$ behaviour of $S_{CC}(q)$ for the BMLJ and BKS system. Now we analyse the difference arising in the kinetic part. In absence of cross correlations, the Onsager coefficient can be estimated from the partial diffusion constants $D_A$ and $D_B$ through the Darken's equation~\cite{darken1948diffusion}:

$$
\Lambda_D = (1 - x) D_A + x D_B.
$$

Therefore, the importance of cross correlations can be quantified by looking at the ratio between the interdiffusion constant and the Darken prediction. The resulting quantity, called the Manning factor in the context of chemical diffusion in crystals~\cite{manning1961diffusion}, is defined as

$$
S \equiv \frac{\Lambda}{\Lambda_D}.
$$

 \begin{table}[h!]
 \centering
 \begin{tabular}{|l|c|c|c|c|c|c|}
 \hline
 System & $D_A$ & $D_B$ & $\Lambda$ & $\Lambda_D$ & $S$\\
 \hline
BMLJ & $5.83 \cdot 10^{-5}$ & $1.7 \cdot 10^{-4}$ & $1.4 \cdot 10^{-4}$ & $1.48 \cdot 10^{-4}$ & 0.95\\
BKS 2.36 g/cm$^3$ & $1.17 \cdot 10^{-4}$ & $1.80 \cdot 10^{-4}$ & $4.17 \cdot 10^{-5}$ & $1.38 \cdot 10^{-4}$ & 0.33\\
BKS 2.48 g/cm$^3$ & $2.07 \cdot 10^{-4}$ & $3.08 \cdot 10^{-4}$ & $1.07 \cdot 10^{-4}$ & $2.41 \cdot 10^{-4}$ & 0.44\\
BKS 2.60 g/cm$^3$ & $3.54 \cdot 10^{-4}$ & $5.05 \cdot 10^{-4}$ & $1.99 \cdot 10^{-5}$ & $4.04 \cdot 10^{-4}$ & 0.49\\
 \hline
 \end{tabular}
 \caption{Values for quantities related to concentration fluctuations and interdiffusion for the BMLJ and BKS systems. $D_A$, $D_B$, $\Lambda$ and $\Lambda_D$ are expressed in MD units for the BMLJ system and nm$^2$/ps for BKS, whereas $S$ is a dimensionless number.}
 \label{tbl:interdiffusion}
 \end{table}
 
Table~\ref{tbl:interdiffusion} shows the values for $D_A$, $D_B$, $\Lambda$, $\Lambda_D$ and $S$ for all the investigated binary mixtures. For the BMLJ system $S \approx 1$, which is a sign that cross correlations are very weak. By contrast, all BKS systems exhibit values of $S$ which differ significantly from $1$, demonstrating the importance of cross correlations also in the interdiffusion kinetic factor.

\end{document}